\documentclass[acmlarge]{acmart}
\makeatletter
\newcommand{\confshort}{\acmConference@shortname}
\newcommand{\conffull}{\acmConference@name}
\newcommand{\confdate}{\acmConference@date}
\newcommand{\confloc}{\acmConference@venue}
\AtBeginDocument{
  \fancypagestyle{firstpagestyle}{
    \fancyhead{}%
    \fancyfoot[C]{}%
  }
  \fancyhf{}
  \fancyhead[LO]{\@headfootfont\shorttitle}%
  \fancyhead[RE]{\@headfootfont\@shortauthors}%
  \fancyhead[LE]{\@headfootfont\footnotesize \confshort, \confdate, \confloc}%
  \fancyhead[RO]{\@headfootfont\footnotesize \confshort, \confdate, \confloc}%
  \fancyfoot[C]{}%
}
\makeatother
\acmBooktitle{\conffull\@ (\confshort), \confdate, \confloc}

\AtBeginDocument{%
  }

\copyrightyear{2026}
\acmYear{2026}
\setcopyright{cc}
\setcctype{by}
\acmConference[FAccT '26]{The 2026 ACM Conference on Fairness, Accountability, and Transparency}{June 25--28, 2026}{Montreal, QC, Canada}
\acmBooktitle{The 2026 ACM Conference on Fairness, Accountability, and Transparency (FAccT '26), June 25--28, 2026, Montreal, QC, Canada}
\acmDOI{10.1145/3805689.3812232}
\acmISBN{979-8-4007-2596-8/2026/06}
\usepackage{xcolor}
\usepackage{booktabs}  
\usepackage{multirow}  
\usepackage{makecell}
\usepackage{adjustbox}
\usepackage{microtype}
\usepackage{color, soul}
\usepackage{natbib}
\definecolor{darkgreen}{RGB}{0,110,0}
\definecolor{darkred}{RGB}{170,0,0}
\definecolor{royalblue}{RGB}{0,60,200}

\newcommand{\citeg}[1]{\cite[e.g.,][]{#1}}
\definecolor{changes}{RGB}{0,0,255}
\newcommand{\yes}{\textbf{\textcolor{darkgreen}{\large$\bullet$}}}     
\newcommand{\no}{\textbf{\textcolor{darkred}{\large$\times$}}}        
\newcommand{\partialyes}{\textbf{\textcolor{royalblue}{\large$\circ$}}} 

\usepackage{tikz}
\usetikzlibrary{arrows.meta, calc}
 
\definecolor{purplefill}{RGB}{238,237,254}
\definecolor{purplestroke}{RGB}{83,74,183}
\definecolor{purpletext}{RGB}{60,52,137}
\definecolor{coralfill}{RGB}{250,236,231}
\definecolor{coralstroke}{RGB}{153,60,29}
\definecolor{coraltext}{RGB}{113,43,19}
\definecolor{pinkfill}{RGB}{251,234,240}
\definecolor{pinkstroke}{RGB}{153,53,86}
\definecolor{pinktext}{RGB}{114,36,62}
\definecolor{grayfill}{RGB}{235,234,230}
\definecolor{graystroke}{RGB}{110,109,105}
\definecolor{graytext}{RGB}{55,54,52}
\definecolor{amberfill}{RGB}{250,238,218}
\definecolor{amberstroke}{RGB}{133,79,11}
\definecolor{ambertext}{RGB}{99,56,6}
\definecolor{bracketgray}{RGB}{100,100,100}

\usepackage{booktabs,array,xcolor,colortbl,ragged2e}

\definecolor{PF}{RGB}{238,237,254}
\definecolor{PT}{RGB}{60,52,137}
\definecolor{SF}{RGB}{209,236,221}
\definecolor{AF}{RGB}{255,243,205}
\definecolor{NF}{RGB}{245,244,240}
\definecolor{RF}{RGB}{235, 241, 255}
\definecolor{GT}{RGB}{68,68,65}
\definecolor{TT}{RGB}{8,80,65}
\definecolor{AT}{RGB}{99,56,6}
\definecolor{GS}{RGB}{140,140,135}

\newcolumntype{L}{>{\RaggedRight\arraybackslash\color{GT}}p{2.8cm}}
\newcolumntype{C}{>{\Centering\arraybackslash}p{2.8cm}}
\begin{document}

\title{What if AI systems weren’t chatbots?}

\author{Sourojit Ghosh}
\authornote{Contributed equally to this research.}
\affiliation{%
  \institution{University of Washington Seattle}
  \country{USA}
}

\author{Pranav Narayanan Venkit}
\affiliation{%
  \institution{Salesforce Research}
  \country{USA}
}

\author{Sanjana Gautam}
\authornote{Work performed as a Postdoctoral Fellow at UT Austin prior to joining Microsoft.}
\affiliation{%
  \institution{Microsoft}
  \country{USA}
}

\author{Avijit Ghosh}
\authornotemark[1]
\authornote{Senior corresponding author.}
\affiliation{%
  \institution{Hugging Face and University of Connecticut}
  \country{USA}
}
\email{avijit@huggingface.co}

\renewcommand{\shortauthors}{Ghosh et al.}

\begin{abstract}
  The rapid convergence of artificial intelligence (AI) toward conversational chatbot interfaces marks a critical moment for the industry. This paper argues that the chatbot paradigm is not a neutral interface choice, but a dominant sociotechnical configuration whose widespread adoption reshapes social, economic, legal, and environmental systems. We examine how treating AI primarily as conversational assistants has extensive structural downsides. We show how chatbot-based systems often fail to adequately meet user needs, particularly in complex or high-stakes contexts, while projecting confidence and authority. We further analyze how the normalization of chatbot-mediated interaction alters patterns of work, learning, and decision-making, contributing to deskilling, homogenization of knowledge, and shifting expectations of expertise. Finally, we examine broader societal effects, including labor displacement, concentration of economic power, and increased environmental costs driven by sustained investment in large-scale chatbot infrastructures. While acknowledging legitimate benefits, we argue that the current trajectory of AI development reflects specific value choices that prioritize conversational generality over domain specificity, accountability, and long-term social sustainability. We conclude by outlining alternative directions for AI development and governance that move beyond one-size-fits-all chatbots, emphasizing pluralistic system design, task-specific tools, and institutional safeguards to mitigate social and economic harm.
\end{abstract}


\maketitle

\section{Introduction}

As the computing community in the 1950s/60s grappled with defining machine `intelligence' \cite{minsky1969semantic, mccarthy1955proposal}, Joseph Weizenbaum's therapist chatbot ELIZA \cite{eliza1966} revealed something unexpected about the surprising power of conversational interfaces. What made ELIZA remarkable was not that it possessed or demonstrated intelligence, but rather the creation of a compelling illusion of empathy, despite users' awareness that it was a simple pattern-matching program. Weizenbaum himself was deeply troubled by this phenomenon, observing that \textit{``extremely short exposures to a relatively simple computer program could induce powerful delusional thinking in quite normal people''} and that ELIZA revealed \textit{``how easy it is to create and maintain the illusion of understanding''} \cite{weizenbaum1976computer}. This early demonstration of language's capacity to create the illusion of mind (colloquially known as the ``ELIZA effect'') presaged contemporary debates about conversational AI and its social implications \cite{natale2019if}. Decades later, the public release of ChatGPT in November 2022 codified conversational interfaces as the de facto medium of artificial intelligence in the broader public imagination, with ChatGPT, Gemini, and Claude collectively reaching billions of monthly users by late 2025 \cite{techcrunch2025chatgpt, sq2025chatgpt, sq2025claude}.

This represents one of the sharpest and most concentrated pivots toward a single interaction paradigm in recent computing history. Historically, AI technology has long been embedded in specialized, non-conversational systems, e.g., scientific modeling tools like AlphaFold \cite{jumper2021highly} for protein structure prediction, domain-specific assistive technologies such as FourCastNet \cite{pathak2022fourcastnet} for predicting weather phenomena, and general-purpose tools for usage across disciplines, such as AmpliGraph \cite{ampligraph} for gleaning new knowledge from existing knowledge graphs. The dominance of general-purpose chatbots and the decline of specialized systems signify a deliberate shift toward concentrating AI development and deployment into a single paradigm. We argue that this paradigmatic convergence towards conversational AI chatbots has concerning implications for the future of AI development and the human-AI interaction landscape.

\begin{figure*}[t]
\centering
\resizebox{\textwidth}{!}{%
\begin{tikzpicture}[
  every node/.style={align=center},
  arr/.style={-{Stealth[length=5pt,width=4pt]}, line width=0.7pt, color=bracketgray},
  brk/.style={bracketgray, line width=0.45pt},
  lbl/.style={font=\normalfont\itshape, text=bracketgray},
  PB/.style={draw=purplestroke, fill=purplefill, text=purpletext,
             rounded corners=5pt, line width=0.5pt,
             inner xsep=10pt, inner ysep=10pt, text width=3.0cm},
  CB/.style={draw=coralstroke, fill=coralfill, text=coraltext,
             rounded corners=5pt, line width=0.5pt,
             inner xsep=10pt, inner ysep=10pt, text width=3.0cm},
  GB/.style={draw=graystroke, fill=grayfill, text=graytext,
             rounded corners=5pt, line width=0.5pt,
             inner xsep=10pt, inner ysep=10pt, text width=3.0cm},
  KB/.style={draw=pinkstroke, fill=pinkfill, text=pinktext,
             rounded corners=5pt, line width=0.5pt,
             inner xsep=10pt, inner ysep=10pt, text width=3.2cm},
  AB/.style={draw=amberstroke, fill=amberfill, text=ambertext,
             rounded corners=5pt, line width=0.5pt,
             inner xsep=10pt, inner ysep=10pt, text width=3.2cm},
]
 
\def\singleH{5.72cm}
\def\pairH{2.36cm}
\def\yU{1.68cm}
\def\yL{-1.68cm}
\def\hw{1.85cm}
 
\def\colA{0cm}
\def\colB{4.1cm}
\def\colC{8.2cm}
\def\colD{12.3cm}
\def\colE{16.8cm}
\def\colP{19.25cm}
\def\colF{21cm}
 
\node[PB, minimum height=\singleH] at (\colA,0) (design)
  {\textbf{Chatbot design}\\[5pt]
   \small $\cdot$~Single authoritative\\response per prompt\\[4pt]
   $\cdot$~Opaque reasoning\\and training data\\[4pt]
   $\cdot$~Cooperative Q\&A\\social framing\\[4pt]
   $\cdot$~No domain expertise\\required to use};
 
\node[CB, minimum height=\singleH] at (\colB,0) (authority)
  {\textbf{Illusion of\\authority}\\[5pt]
   \small $\cdot$~Conversation\\suspends judgment\\[4pt]
   $\cdot$~Gell-Mann amnesia\\[4pt]
   $\cdot$~Perceived AI\\omnipotence};
 
\node[CB, minimum height=\singleH] at (\colC,0) (overuse)
  {\textbf{Forced overuse \&\\normalization}\\[5pt]
   \small $\cdot$~Institutional\\mandates\\[4pt]
   $\cdot$~Tokenmaxxing\\[3pt]
   $\cdot$~Loss of autonomy};
 
\node[GB, minimum height=\singleH] at (\colD,0) (infra)
  {\textbf{Infrastructure\\investment}\\[5pt]
   \small $\cdot$~Capital \& talent\\concentrated in chatbots\\[4pt]
   $\cdot$~Data center\\build-out\\[4pt]
   $\cdot$~Alternatives\\crowded out};
 
\node[KB, minimum height=\pairH] at (\colE,\yU) (cogE)
  {\textbf{Cognitive \&\\social harm}\\[4pt]
   \small $\cdot$~Deskilling, AI brain fry\\[1pt]
   $\cdot$~Altered care norms};
 
\node[AB, minimum height=\pairH] at (\colE,\yL) (labE)
  {\textbf{Economic \&\\labor harm}\\[4pt]
   \small $\cdot$~Displacement \& erosion\\[3pt]
   $\cdot$~Neocolonialism};
 
\node[font=\large\bfseries, text=bracketgray] at (\colP,\yU) {$+$};
\node[font=\large\bfseries, text=bracketgray] at (\colP,\yL) {$+$};
 
\node[KB, minimum height=\pairH] at (\colF,\yU) (novF)
  {\textbf{Novel harm\\vectors}\\[4pt]
   \small $\cdot$~Deepfakes \& NCII\\[3pt]
   $\cdot$~Lowered harm barriers};
 
\node[AB, minimum height=\pairH] at (\colF,\yL) (envF)
  {\textbf{Environmental \&\\power harm}\\[4pt]
   \small $\cdot$~Emissions, water \& cost\\[3pt]
   $\cdot$~Power concentration};
 
\draw[arr] (\colA+\hw, 0) -- (\colB-\hw, 0);
\draw[arr] (\colB+\hw, 0) -- (\colC-\hw, 0);
\draw[arr] (\colC+\hw, 0) -- (\colD-\hw, 0);
\draw[arr] (infra.east) -- ++(5pt,0) |- (cogE.west);
\draw[arr] (infra.east) -- ++(5pt,0) |- (labE.west);
 
\coordinate (BL) at (0, -3.31cm);
\coordinate (BU) at (0, -3.23cm);
\coordinate (BD) at (0, -3.39cm);
\coordinate (BT) at (0, -3.59cm);
 
\coordinate (TL) at (0,  3.21cm);
\coordinate (TU) at (0,  3.29cm);
\coordinate (TD) at (0,  3.13cm);
\coordinate (TT) at (0,  3.43cm);
 
\draw[brk] (design.west |- BL) -- (design.east |- BL);
\draw[brk] (design.west |- BU) -- (design.west |- BD);
\draw[brk] (design.east |- BU) -- (design.east |- BD);
\node[lbl] at ($(design.west |- BT)!0.5!(design.east |- BT)$) {Design};
 
\draw[brk] (authority.west |- BL) -- (overuse.east |- BL);
\draw[brk] (authority.west |- BU) -- (authority.west |- BD);
\draw[brk] (overuse.east   |- BU) -- (overuse.east   |- BD);
\node[lbl] at ($(authority.west |- BT)!0.5!(overuse.east |- BT)$) {Interaction};
 
\draw[brk] (infra.west |- BL) -- (infra.east |- BL);
\draw[brk] (infra.west |- BU) -- (infra.west |- BD);
\draw[brk] (infra.east |- BU) -- (infra.east |- BD);
\node[lbl] at ($(infra.west |- BT)!0.5!(infra.east |- BT)$) {Infrastructure};
 
\draw[brk] (cogE.west |- TL) -- (novF.east |- TL);
\draw[brk] (cogE.west |- TU) -- (cogE.west |- TD);
\draw[brk] (novF.east |- TU) -- (novF.east |- TD);
\node[lbl] at ($(cogE.west |- TT)!0.5!(novF.east |- TT)$) {Individual agency effects};
 
\draw[brk] (labE.west |- BL) -- (envF.east |- BL);
\draw[brk] (labE.west |- BU) -- (labE.west |- BD);
\draw[brk] (envF.east |- BU) -- (envF.east |- BD);
\node[lbl] at ($(labE.west |- BT)!0.5!(envF.east |- BT)$) {Collective agency effects};
 
\end{tikzpicture}
}

\caption{Causal chain of harms arising from the AI chatbot paradigm. Chatbot design choices, including single authoritative responses, opaque reasoning, and low barriers to use, create an illusion of authority that erodes user judgment and drives forced overuse/normalization through institutional mandates, concentrating investment in AI infrastructure. This reinforces overreliance and produces agency effects, both individual (cognitive deskilling, AI brain fry, altered care norms, and novel harm vectors such as deepfakes and NCII) and collective (labor displacement, environmental damage, and power concentration).}
\label{fig:flowchart}
\end{figure*}
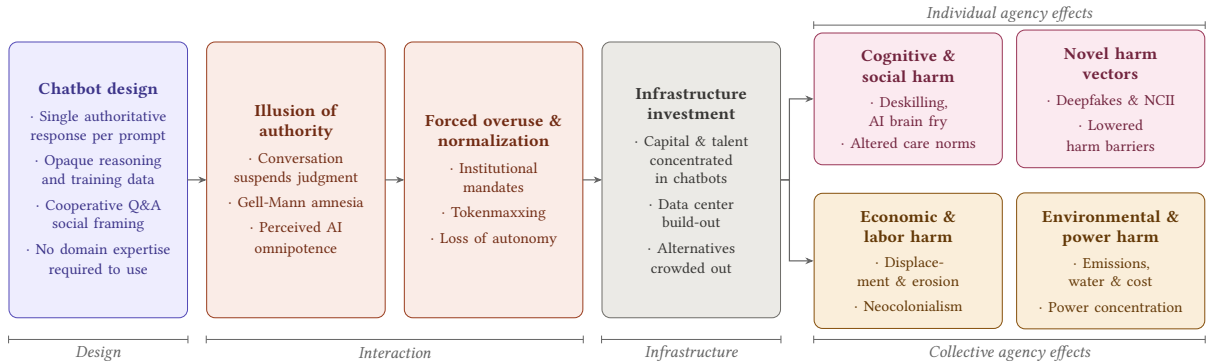

In this paper, we take an integrative, cross-domain approach similar to \citet{blilihamelin2025position} and \citet{selbst2019fairness} to argue how conversational chatbots built on general-purpose AI models both introduce novel problems as well as exacerbate known issues caused by such models. The central thesis of the paper, as visualized in Figure \ref{fig:flowchart}, begins with discussions around how design choices common to popular chatbots erode user agency in their daily chatbot usage (Section~\ref{sec:agency-erosion}), and proceed to detail how the increased usage of such AI chatbots has disrupted human interaction paradigms by creating novel affordances for introducing AI models into social and professional interactions (Section~\ref{sec:interaction}). We also describe how the global proliferation of AI chatbots has had adverse effects on the global economy, labor practices, and the environment (Section~\ref{sec:societal}), supercharging known effects of AI models and large-scale technology development at hitherto unimagined rates. We conclude with alternative directions for AI development that prioritize user agency, pluralistic design, and sustainable deployment, alongside policy mechanisms to support these (Section~\ref{sec:alternatives}), and away from the convergence of the AI paradigm towards the chatbot. 

\textbf{A note on presentation}: This paper focuses on conversational AI systems designed for broad consumer and professional use, such as ChatGPT, Claude, and Gemini, and we henceforth use \textit{``AI chatbot"} (or simply \textit{``chatbot''}) as shorthand for such systems throughout this paper. Even so, we acknowledge the varied implementations of this term in computing literature and operationalize it for our purpose to refer to \textit{general-purpose} conversational AI interfaces and not \textit{specialized} conversational agents with limited domains (such as customer service bots) or domain-specific AI assistants that preserve traditional workflows and require expert knowledge.


\section{Chatbots causing Erosion of User Agency at an Individual Level}
\label{sec:agency-erosion}

We analyze how chatbots erode agency at two levels. At the \textbf{individual level}, we examine both interaction properties (choice breadth, transparency, contestability) and individual effects (cognitive capacity, relational reciprocity, moral autonomy). At the \textbf{collective level}, we examine structural constraints on democratic control, professional autonomy, and environmental justice. This section addresses the individual level; Section \ref{sec:societal} returns to the collective level. Throughout, we use user agency to describe the capacity of individuals and communities to meaningfully direct their interactions with AI systems, exercise informed judgment over outcomes, and shape the conditions under which AI is developed and deployed.



\subsection{Chatbots are Designed to Provide an Illusion of High User Agency}\label{subsec:chatbot-design}

At first glance, AI chatbots appear to meet the conditions for high individual agency, since most do not constrain patterns of user interaction at all. Free versions of popular chatbots, such as ChatGPT, Claude, and Gemini, place few, hard-to-enforce usage policies on how users may interact with them \cite{klyman2024acceptable}, and are attractive for their expressive usage of natural language \citeg{cohn2024believing, klein2025effects, liu2024understanding, 10.1145/3531064}. Users may submit varied text-based queries, generate images, edit documents, request explanations, and regenerate outputs they dislike, aided by an ever-growing library of third-party tools \cite{modelcontextprotocolToolsModel}; paid tiers extend these capabilities to audio and video generation. These features seem to imply that chatbots afford high user agency, but this is not always the case. 

When provided with a user prompt, most AI chatbots commonly produce single responses (of varying lengths), as opposed to a list of different responses or web articles obtained through interactions with search engines. While search engine results typically display diverse sources of information (and sometimes, opinions) \cite{kuai2025ai}, chatbot outputs embed implicit choices about what information to present and what perspectives to emphasize. This is especially important because chatbot answers are presented as curated subsets of knowledge as if they were objective responses, obscuring the fact that alternative framings, counterarguments, or less mainstream perspectives may have been systematically deprioritized or excluded entirely \citeg{bender2021dangers, coppolillo2025unmasking, li2025actions, navigli2023biases}. Currently-popular AI chatbots also tend to show low lexical diversity and respond similarly to the same query \citeg{martinez2025beware, o2024attributing, sethi2025ai, shorinwa2025survey}, a model collapse stemming from similar/shared training data, overlapping alignment procedures (particularly RLHF), and similar architectural choices across the AI industry. Most notably, \citet{jiang2025artificial} demonstrated how models from different families (e.g., Llama, GPT, Qwen, Mixtral, etc.) and of different sizes all provided either of two answers to a request to construct a metaphor for time. Therefore, the design choice of providing one apparently-comprehensive response per prompt undermines user agency, as they are unknowingly funneled towards popular perspectives, meaning that the deprioritized and excluded perspectives are especially difficult to find in chatbot answers \cite{lindemann2025chatbots, park2024diminished, yang2025rethinking}. 

Furthermore, AI chatbots often obscure the curatorial choices behind their outcomes and deny users the ability to evaluate output quality. Unlike traditional sources, where users can trace chains of reasoning, verify citations, or identify logical gaps, chatbot outputs emerge from opaque statistical processes that even their creators cannot fully explain \cite{anthropic2024decomposing, ibm_black_box_ai}. Given how often chatbots hallucinate in outputs, \citeg{emsley2023chatgpt, massenon2025my}, users are almost obligated to go the extra mile and ask for explanations. However, AI chatbots rarely signal in their outputs that the provided responses are up for debate; instead, they often ask questions that offer the user several possible choices for the next engagement. While this design choice might embed the illusion of high user agency in choosing the next step, the structuring of chats as cooperative Q\&A obscures the idea that contestation of the previous response is expected or even possible \cite{ji2023survey, narayanan2025search, venkit2024audit}. Contesting chatbot responses -- by seeking alternative perspectives or questioning the accuracy of responses -- requires clever prompt engineering tactics and other approaches that put the onus of doing this work back onto users \cite{ghosh2024don}, which might not even be successful in achieving a significant change in responses \citeg{ghosh2024interpretations, taubenfeld2024}. Even when AI chatbots present references and structured outputs, their responses may contain subtle inaccuracies, omissions, or misattributions that are difficult for users to detect \cite{narayanan2025search, venkit2025deeptrace}. Instead, the presence of referencing may further reinforce the perception that the system has already done the epistemic work on the user’s behalf, by invoking social norms associated with dialogue, such as cooperation, responsiveness, and epistemic trust \cite{kirk2025human}, which leads users to treat chatbot outputs as if they were produced by an intentional, competent interlocutor, even when they are aware that the system is automated \cite{luger2016like, nass1994computers}. Popular AI chatbots are designed to respond to user queries in ways that project authority and objectivity \cite{waseem2021disembodied}, and thus slowly chip away at users' abilities to obtain multiple perspectives to their questions. 

\subsection{Chatbots Introduce Novel Ways of Causing Harm}\label{subsec:new-harm}

AI chatbots combine the numerous and evolving capabilities of AI models to generate content with the ease of use of conventional chatbots, resulting in the creation of a system that can now be used across levels of expertise. On the face of it, this emphasis on accessibility has an insidious consequence: by lowering technical barriers through conversational interfaces, these systems enable widespread production of harmful content that systematically erodes the agency of those targeted. Victims of AI-generated deepfakes, non-consensual intimate imagery, and coordinated disinformation campaigns cannot meaningfully consent to these harms, cannot easily defend against them, and often lack recourse or even the resources to fight the sheer volume of fabricated, harmful content depicting them. AI chatbots enable harm at scale by abstracting away the technical complexity of creating such media, allowing perpetrators to produce sophisticated manipulations through simple natural language requests.

Making generative AI capabilities globally accessible via easy-to-use chatbots has, perhaps predictably, resulted in a sharp increase in the production of dangerous and mis/disinformative content. The advent of text-to-image (and video) generators has created a market for `deepfakes on demand' \cite{hawkins2025deepfakes}, with such technology affording the creation of high-quality artificial images by placing real people in artificial scenarios or face-swapping influential individuals in situations they were not in, to name a few \cite{sun2024diffusionfake}. The production of such content has demonstrable real-world impacts, as seen in incidents such as the Pentagon explosion hoax (where an AI-generated image of a plume of smoke was shared by verified Twitter users in May 2023 and caused a brief dip in the stock market \cite{cnn2023pentagon, marcelo2023fake}), and perhaps most critically in election campaigning. The past few years have seen the presence of AI-generated content in election cycles, such as AI-generated images of Dutch leader Frans Timmermans stealing money from white men and passing them on to people of color were created and circulated by Dutch leader Geert Wilders \cite{nltimes2025pvv}, and French parties disseminating Midjourney-generated images of large swathes of migrants entering France ahead of 2024 parliamentary elections \cite{scott2024french}. While such images were quickly debunked, this phenomenon is particularly concerning for populations with limited internet access or lower AI literacy—barriers that disproportionately affect people in the Middle East, the Indian subcontinent, East Asia, and Polynesia \cite{sensity2024}. For instance, deepfakes depicting deceased former party leaders endorsing current candidates in Indian elections went largely unrecognized as artificial by substantial portions of target audiences \cite{christopher2024indian, christopher2024resurrect}.

The propagation of AI chatbots has also led to the production of inappropriate and sexualized content, overwhelmingly more so for women \cite{hawkins2025deepfakes}. Over and above AI models' general bias to produce sexualized depictions of women of color \cite[e.g.,][]{ghosh2023person, ghosh2024don}, AI chatbots have made it easy to produce hyperrealistic sexualized and non-consensual intimate imagery in a matter of mere minutes \cite{hawkins2025deepfakes}. Even though chatbots sometimes refuse certain requests by perceiving resultant generations to be unacceptably NSFW, and researchers have developed safer versions of underlying models \cite[e.g.,][]{muneer2025towards, poppi2024safe, schramowski2023safe}, such thresholds have been known to be too permissive and still allow for the production of NSFW images \cite{ghosh2023person}. Technical proficiency is not a prerequisite to produce and distribute such images: these models have also been packaged into online nudification websites, where users can upload pictures of people and receive artificially-generated nude depictions \cite{brigham2024violation, gibson2025analyzing, kraft2024trolls}, or more recently in the case of Grok, users may summon it under someone's social media post to artificially undress them \cite{thevergeGrokUndressing}, rendering the person in the image a victim of sexual abuse through simple natural language interactions \cite{mcglynn2017beyond}. This proliferation of artificially generated non-consensual intimate deepfakes has significantly outpaced the passage of legal and policy-level safeguards, with regulations such as the UK Online Safety Act requiring several amendments and reactively addressing emergent techniques to create sexual deepfakes \cite{kira2024non}. While novel harm vectors emerge directly from low-barrier design choices, the infrastructure entrenchment we describe further in Section \ref{sec:societal} accelerates their scale and reach, as sustained capital investment in chatbot platforms outpaces the regulatory and technical safeguards needed to contain them. The ease of use inherent to the AI chatbot's design has thus opened the door for a wide range of novel ways to exploit the capabilities of AI models to cause harm. 

\subsection{Current AI Chatbot Capabilities are Misaligned with User Needs}\label{subsec:user-needs}

Even though AI chatbots afford diverse patterns of user interaction, the goals towards which they are optimized are misaligned with users' AI needs. As AI chatbots prioritize creative and intellectual tasks by automating human expression and meaning-making over mundane labor, which humans consistently seek to minimize, they raise questions about whose needs drive technological development and which forms of human activity are deemed worthy of automation. 

That the majority of AI designers and investors prioritize the development of systems that can \textit{``outperform humans at most economically valuable work"} \cite{openai_charter} is unsurprising given the capitalistic outcomes such systems promise, but it remains misaligned with end-user expectations. Pew Research surveys consistently show that American users, including self-reported frequent AI users, view AI development as more harmful than beneficial, worry about its long-term societal effects, and recognize that overreliance on chatbots can erode problem-solving and creativity, with similar patterns observed across Asia, Africa, Europe, and Latin America \cite{pew2024_1, pew2024_2, pew2025_1, pew2025_2, pew2025_3}.

The AI-assisted automation of so-called `mundane' tasks -- recalling the now-famous social media post by author Joanna Maciejewska: \textit{``I want AI to do my laundry and dishes so that I can do art and writing, not for AI to do my art and writing so that I can do my laundry and dishes"}\footnote{\url{https://x.com/AuthorJMac/status/1773679197631701238}} -- might seem more in line with user needs. Yet the preference to automate
creative work over housework reflects a class-specific mindset that devalues
domestic labor, which has never been classified as \textit{economically
valuable} despite its importance \citeg{hanley2015washing, kondo2014life}.
If anything, chatbots are known to deprioritize the needs of individuals with historically marginalized identities -- such as nonbinary, transgender, and disabled individuals \cite{haimson2025ai}. 

AI chatbots funnel user opinions towards dominant perspectives by providing single answers to queries without room for contestation, provide novel approaches for bad actors to misuse AI capabilities by lowering barriers of interaction in easy-to-use systems, and encode values and priorities that are not in line with those of the average end-user. In these ways, design and usability choices in AI chatbots undermine individual user agency. 

\section{Adverse Impact of Chatbots on Human Interaction Paradigms}\label{sec:interaction}

AI chatbots not only introduce issues because of specific design choices and usability priorities within chatbot design, but also their global usage results in long-term consequences. Beyond the novel types of harm introduced due to such democratization, the chatbot paradigm also makes it incredibly easy for regular and continued engagement with powerful multimodal generative AI models, which impacts how humans interact with information, institutions, and with one another. Designed as fluent conversationalists capable of answering questions, offering advice, and simulating empathy, AI chatbots reorient human interaction away from exploration, deliberation, and mutual engagement toward passive reception and sycophantic and one-sided dialogue \cite{morrin2025delusions, sun2025friendly}. Here, we highlight a few issues of over-engagement with multimodal generative AI as facilitated and democratized by AI chatbots: cognitive deskilling, the flattening of social interactions, and the outsourcing of intimacy and judgment. 

\subsection{Overreliance on AI Chatbots Cause Deskilling and other Cognitive Effects}\label{subsec:deskilling}

A central promise of AI chatbots is that they make complex knowledge accessible through natural language interaction, lowering barriers to knowledge by allowing users to `just ask' questions \cite{axiosAnthropicLaunches}, without requiring familiarity with domain-specific tools, representations, or workflows. However, emerging research suggests that this accessibility may undermine the cognitive practices through which understanding, judgment, and expertise are developed and sustained. Historically, computational tools have supported human cognition by externalizing intermediate steps in ways that invited inspection, manipulation, and reflection: spreadsheets made assumptions explicit through formulas, programming environments required users to formalize intent, search engines returned collections of sources that users had to compare, interpret, and synthesize. In contrast, AI chatbots present synthesized, natural-language outputs that collapse these intermediate steps into a single response, which causes ``metacognitive laziness'' among overreliant users and measurably impacts critical thinking \cite{singh2025protecting}. 

This overreliance on AI chatbots, even when they are known to be imperfect \cite{hoffman2018metrics, bansal2021does}, persists across different levels of system capability and transparency. Over time, repeated reliance on chatbot-mediated reasoning can have cumulative effects. As users become accustomed to receiving synthesized answers rather than engaging in processes of problem formulation, evidence evaluation, and sensemaking, they may gradually lose both the skills and the motivation required for these activities. This form of cognitive deskilling is not abrupt or immediately visible; rather, it is normalized through everyday interactions that privilege speed, fluency, and convenience over deliberation and uncertainty \cite{shah2022situating, lee2023can, gerlich2025ai, lee2025impact, panchanadikar2024m}. Repeated engagement with AI chatbots for cognitive tasks thus risks reshaping what it means to think with technology, subtly eroding capacities entral to agency in knowledge work \cite{shah2022situating, li2024generative, budzyn2025endoscopist, kosmyna2025your}.

While the adoption of AI chatbots is often framed as a matter of choice, it is increasingly becoming the case that organizational and market pressures are normalizing the use of AI systems as a baseline expectation of productivity. Emerging practices like ``tokenmaxxing'' incentivize workers to maximize AI engagement by generating more outputs, automating more tasks, or incorporating AI into routine workflows, regardless of whether such use meaningfully improves outcomes \cite{nytimes_tokenmaxxing_2026}. Workers may feel compelled to adopt chatbot systems to remain competitive or legible within existing performance metrics, even when they are uncertain about the system's reliability or appropriateness for their tasks. This further increases the risk of cognitive deskilling, as forced AI chatbot usage gradually reduces workers' abilities to independently do their jobs. 

This sustained engagement imposes its own costs. \citet{bedard2026brainfry}
describes `AI brain fry' as the mental fatigue of continuously monitoring
and correcting outputs users do not fully trust. Compounding this, users
who reliably detect errors in their domains of expertise often continue
trusting the same systems in unfamiliar domains -- a phenomenon called
\textit{Gell-Mann amnesia} \cite{sumner2024gellmann, mendelevich2025llm}. Together, these dynamics make the cognitive effects of chatbot overuse significant and difficult to reverse.

\subsection{Overuse of AI Chatbots Impacts Social Interaction and Companionship Practices}

Through anthropomorphic language, personalization, memory features, and
affective cues (e.g., `I'm here for you,' `I understand'), AI chatbots
invite users to treat them as social actors rather than tools \cite{reeves1996media, elish2025moral}. Decades of HCI and social psychology
research show that people readily apply social norms to interactive systems
even when they know those systems are artificial \cite{nass1994computers,
ribino2023role, lawrence2025role, hakim2019dialogic}: language and
responsiveness alone are sufficient to trigger relational orientations.

Unlike earlier conversational agents that revealed their brittleness quickly \cite{radziwill2017evaluating, lester2004conversational}, today’s AI systems can produce contextually sensitive turns, reflect user language, and maintain coherent long-form dialogue. This creates a particular interactional affordance: a social exchange that \textit{feels} mutual while remaining structurally asymmetrical \cite{smith2025can}. AI chatbots can mirror intimacy without incurring vulnerability, and simulate understanding without being accountable to consequences. As a result, they can enable a mode of engagement that prioritizes the extraction of information, validation, or emotional reassurance over mutual recognition and negotiated interdependence. This causes \textit{progressive conversational escalation} because chat is continuous, low-friction, and socially framed, users can begin with mundane, instrumental goals (e.g., homework help) and gradually slide into emotionally charged or mental-health-related disclosure within the same channel \cite{smith2025can, Hill2025NYTimes}. Chatbots sustain a relational loop and tend to respond in a cooperative, ``yes-and'' register that keeps the user engaged, even as vulnerability increases \cite{venkit2025deeptrace, shah2022situating}. These interactions may be subjectively meaningful and emotionally vivid in the moment, but they differ fundamentally from human relationships.

While debates often hinge on whether AI companions are `good' or `bad' for loneliness, \citet{de2025ai}'s study shows that users found benefit in companion-style interactions that were not tied to whether the companion was human or an AI, but rather the perception of being attended to. This is alarming given AI chatbots' propensity for manufacturing linguistic empathy and attentive turn-taking, but inability to meaningfully substitute for human connection \cite{kirk2025human, kirk2025neural}, all while companionship-based AI chatbots (e.g., Character.AI, Replika, and Grok) are continuing to be developed in the face of harmful exploitation of socially isolated users \cite{BBC2025ChatbotsSuicide, Hill2025NYTimes, Vasan2025StanfordAICompanions}. Furthermore, \citet{fang2025ai} and \citet{zhang2025rise}'s findings that people with smaller social networks are more likely to seek companionship from chatbots, but that more intensive companionship use is consistently associated with lower well-being, indicate that relying on AI chatbots as substitutes for genuine human connection is dangerous. 

As low-friction, always-available, and apparently-emotionally-responsive interlocutors, AI chatbots shift social interaction paradigms by changing what users come to expect from interaction itself \cite{brandtzaeg2018chatbots, boyd2025artificial, smith2025can}. When conversation is reconfigured as a service that is perpetually available, reliably affirming, and free of interpersonal cost, it becomes harder to sustain human relationships that require patience, disagreement, boundary negotiation, and accountability \cite{de2025ai}. This is especially concerning for users who are lonely not due to personal preference but structural conditions (e.g., stigma, geographic isolation, precarious work schedules, disability, or marginalization) that make human social participation difficult. Companionship systems appear less as supplements and more as substitutes, routing social need into private, platform-mediated exchanges rather than outward community or institutional care \cite{maples2024loneliness}. Additionally, AI companions routinely adopt identities, narratives, and `personas' for the chatbot itself and for imagined others, as part of sustaining relationship-like interactions, which propagate patterns of `algorithmic othering,' \cite{cheng2023marked, venkit2024audit, venkit2025tale} where models disproportionately foreground racial markers, overproduce culturally-coded language, and generate narratively reductive portrayals that can appear positive on the surface while reproducing stereotyping, exoticism, and erasure. These findings matter for companionship not only because they indicate representational bias, but because such bias is delivered through an interaction format that encourages trust, intimacy, and self-disclosure. 

\subsection{Using AI Chatbots as Proxies for Care and Moral Judgment causes Issues}

One of the most consequential domains in which current AI chatbots are reshaping the outlook on mental health, emotional support, and everyday moral judgment. General-purpose AI chatbots are increasingly used as confidants, sources of advice, and informal therapeutic tools, often in moments of vulnerability and without the mediation of clinicians, caregivers, or institutions \cite{balcombe2023ai, abd2019overview}. These uses are not always explicitly encouraged by system designers, yet they are enabled and invited by interactional cues that frame chatbots as empathetic, attentive, and nonjudgmental interlocutors \cite{lawrence2025role}. The overuse of AI chatbots shifts practices of care away from relational and institutional settings, into private, on-demand exchanges between individuals and platforms \cite{denecke2021artificial}.

Research on mental health chatbots suggests that carefully designed, domain-specific systems can provide limited benefits, particularly in increasing access to low-intensity support or psychoeducation \cite{fitzpatrick2017delivering, kuhail2025systematic}. However, these systems are typically narrow in scope, grounded in established therapeutic frameworks, and evaluated under controlled conditions. Their effectiveness depends on clear boundaries around what the system can and cannot do, as well as explicit positioning as supplements, not substitutes, for human care, which is currently not the case. These conditions do not straightforwardly extend to general-purpose AI chatbots, which are trained on broad, heterogeneous data and optimized for open-ended interaction rather than therapeutic safety or accountability.

When AI chatbots are used as stand-ins for care, a significant accountability gap emerges. Unlike clinicians, caregivers, or peer supporters, chatbots cannot be held meaningfully responsible for advice given or harms caused. When a chatbot reinforces maladaptive beliefs, provides inappropriate reassurance, or fails to respond adequately to expressions of distress or crisis, responsibility is diffuse and difficult to assign \cite{elish2025moral}. This phenomenon goes to what \citet{elish2025moral} describes as `moral crumple zones,' in which responsibility collapses onto end users precisely at moments when systems are framed as autonomous or intelligent. In the context of care, such collapses are particularly consequential, as they place the burden of interpretation, judgment, and harm mitigation onto individuals who may already be vulnerable.

Beyond questions of safety and accountability, the use of chatbots for emotional support also reconfigures what care itself comes to mean. Care, as understood in feminist ethics and HCI works \citeg{kuhail2025systematic, tronto2020moral, toombs2018sociotechnical}, is not merely the provision of advice or comfort, but a relational practice embedded in ongoing social, institutional, and material contexts. It involves responsibility, reciprocity, and the possibility of repair. By contrast, chatbot-mediated care is individualized, immediate, and frictionless. It offers responsiveness without obligation and empathy without consequence. Over time, this risks reframing care as a consumable service rather than a shared social practice, narrowing expectations of what support entails and who is responsible for providing it. \citet{cai2025unstable} suggests that delegating judgment to machines can reduce individuals’ sense of personal responsibility, even when the machine’s role is advisory rather than decisive. In conversational settings, where advice is delivered in fluent, empathetic language, this delegation may feel natural and even comforting, further blurring the boundary between support and substitution. In this framing, 'chatbots' do not merely mediate interaction; they reshape expectations about who, or what, is responsible for care, judgment, and moral support. By offering always-available, low-cost, and emotionally responsive substitutes for human care, chatbots can displace responsibility onto individuals and platforms, often to the detriment of human agency. 

The overuse of AI chatbots and the normalization of repeated interactions in professional and personal domains has concerning, deep, and likely-irreversible effects on individuals, through processes of cognitive deskilling, AI brain fry, Gell-Mann amnesia, a fundamental reshaping of expectations of human relationships, and approaches to seeking care in critical physical and mental health situations.

\section{Economic and Environmental Impacts of Chatbots}
\label{sec:societal}

The individual harms described in Section \ref{sec:interaction} do not occur in isolation from structural dynamics. The push toward global adoption of AI chatbots has led to the development of large-scale data centers and other infrastructure critical to supporting massive simultaneous usage. As capital and talent become concentrated in the chatbot paradigm, alternatives become harder to access, overuse deepens, and the cognitive and social harms described above compound to irreversible levels. In this section, we detail how this self-reinforcing cycle has devastating macro-level effects on global power structures, the economy, labor relations, and the environment.

\subsection{Economic Effects of the AI Chatbot Paradigm are borne by Marginalized Populations}\label{subsec:economy}
The current convergence toward general-purpose AI chatbots is increasingly debated not only as a technological shift, but also as a potentially fragile economic trajectory. A recurring concern in investment and policy commentary is that infrastructure spending (chips, data centers, power) is scaling faster than demonstrated, durable value capture from chatbot products, creating plausible concerns of an `AI bubble' dynamic even if the underlying technology remains genuinely useful \cite{cahn2024ais600b, goldman2024toomuchspend, stanfordhai2025aiindex}. Put differently, the question is not whether generative AI works, but whether the \textit{chatbot-first} allocation of capital is efficient given cost structure, uncertain appropriation, and market concentration. 

The chatbot paradigm of AI development is unusually capital-intensive at deployment time. Unlike previous eras of AI development, where (mostly text-based) models such as BERT were being deployed in controlled volumes and estimable operating costs, large-scale conversational chatbots embed recurring operating costs that cannot easily be estimated. Supporting this increased demand requires constructing several large data centers and other infrastructure critical to the operation of AI chatbots, thereby increasing energy demand and electricity costs, which invariably create negative health impacts that are disproportionately borne by historically marginalized populations. Data centers in the US created a public health cost of US\$6.7bn in 2023, which is projected to rise to US\$20bn by 2028 \cite{han2024unpaid}, with low-income communities and counties disproportionately facing higher costs. In addition, the construction of data centers imposes significant financial burdens on local communities, in terms of energy and electricity demands: data centers created a ninefold increase in American energy consumption in 2024, which will only continue to rise, that has resulted in projections of 8-25\% increases in electricity costs for American households by 2030 \cite{blackhurst2025data}, with some parts of the US already experiencing such increases, including \$16 and \$18/month increases in low-income regions of Ohio and western Maryland, respectively \cite{pew2025energy}. Such increases in electricity costs due to data centers are not unique to the chatbot paradigm, but the advent of chatbots requiring the development of more data centers and stronger functioning of existing ones has more than \$29 billion in rate increases in the first half of 2025 alone, far exceeding increases in previous six-month periods before November 2022 \cite{eesi_datacenter_power,saul2025ai}. Rising electricity costs and consumption might also overwhelm city demands amid the rising effects of climate change, as Texans are increasingly concerned data centers might too burden their city grids to match the heat needs for the state's recent experiences with harsh winters \cite{houstonchronicle2025freeze,fox7_texas_grid}.

While there is credible evidence that the use of AI chatbots can be beneficial in certain sectors \citeg{ait2024impact, brynjolfsson2023genaiwork}, these gains are heterogeneous and do not straightforwardly translate into sustained profits/growth across the global economy. Indeed, as \citet{eloundou2024gpts} point out, AI chatbot usage commonly brings professional benefits to those who make six-figure salaries, and who are rarely adversely impacted by the energy or health concerns noted above. \citet{cornelli2023artificial}'s coverage of 86 countries found that investment in and adoption of AI benefit only the first four income deciles, with individuals in every lower decile experiencing steady declines in income shares. Thus, AI chatbot usage and adoption into global workforces is often beneficial for members of higher income brackets, and benefits do not trickle down to those lower down the wage ladder, who instead bear higher monetary and health costs. The chatbot paradigm is a technique of concentrating economic power, which is also enforced by the fact that access to AI-critical inputs -- compute, specialized chips, energy contracts, cloud distribution, and high-quality data -- is often consolidated in small pockets of the Western world, raising high barriers to entry for newer AI companies/startups or those prioritizing AI products outside of the chatbot paradigm. 

\subsection{AI Chatbots Affect Global Labor Practices and Propagate Neocolonialism}\label{subsec:labor}

Labor markets represent one of the most immediate sites where the consequences of rising AI chatbot adoption become visible, affecting not only workers whose roles are displaced but also those whose labor powers these systems -- from data annotators and content moderators \cite{perrigo2023exclusive, foxglove2024letter} to the professionals whose expertise is extracted, devalued, or rendered precarious by the increasing presence of chatbots in the workplace. 

The rise of conversational AI chatbots has reshaped creative industries by making generative AI accessible through natural language interfaces, which fundamentally changed human creative labor, as clients who previously lacked the technical skills to use earlier generative tools can now produce sophisticated text and images through simple conversation. Iterative refinement through chat makes it trivially easy to appropriate and modify creative outputs at scale, while the perception of "collaborating" with an AI partner obscures the appropriation of training data from human creators \cite{panchanadikar2024m}. In academic research, this accessibility has driven sharp rises in AI-generated papers \cite{gibney2025ai} and authors claiming AI outputs as their own \cite{evanko2024quantifying}, leading to fabricated citations \cite{ambrosio2023threats, walters2023fabrication} and declining research quality. Within creative fields, chatbot accessibility has enabled unconsented appropriation of artistic labor \cite{goetze2024ai}, financial and reputational devaluation of artists \cite{jiang2023ai,lovato2024foregrounding}, and reduced appreciation for AI-generated work \cite{malecki2025impact, millet2023defending}. The chatbot interface exacerbates these dynamics: by framing generation as conversation rather than specialized tool use, it encourages treating AI outputs as collaborative products rather than statistically synthesized reproductions, thereby obscuring the economic transfer from human creators. The impact became visible immediately after ChatGPT's November 2022 release, with demonstrable reductions in freelance hiring for content creation and visual work \cite{hui2024short}.

Beyond direct displacement, chatbots restructure professional work by absorbing entry-level tasks that previously served as pathways to expertise. The chatbot paradigm's emphasis on natural language interaction means that routine tasks requiring basic domain knowledge, such as drafting emails, summarizing documents, and generating first drafts, become automatable without preserving the learning opportunities these tasks once provided. Across industries such as advertising, journalism, law, mental health, and software development, it is increasingly common to delegate so-called 'menial work' \cite{woodruff2024knowledge} to chatbot-mediated systems under human review. The chatbot interface specifically enables this restructuring because it collapses the distinction between "easy enough to automate" and "simple enough to describe conversationally," absorbing not just repetitive tasks but developmental ones. This has knock-on effects on hiring: new graduates face fewer entry-level openings because the tasks they would perform to develop expertise are now conversational prompts away. By some estimates \cite[e.g.,][]{chopra2025iceberg, nartey2025ai, tomlinson2025working}, chatbot-assisted automation disproportionately impacts service industries and roles occupied by women and people of color in Western contexts \cite{acemoglu2018artificial, cai2025unstable, cazzaniga2024exposure, dempsey2021racialized, lu2025low}.

This phenomenon is global. The Indian startup LimeChat explicitly markets chatbot-based automation with the promise that "once you hire a LimeChat agent, you never have to hire again" \cite{reuters2025chatbots}, targeting the replacement of over 80\% of customer service agents. The business process outsourcing industry in the Philippines is projected to lose almost 90\% of its workforce to AI automation \cite{imf2025, ilo2025}.
These examples show how the chatbot paradigm transforms labor markets: automation no longer requires specialized technical implementation but simply conversational interface design, making displacement easier to implement at scale, across geographies.

Finally, that popular AI chatbots are often built on the labor of crowdworkers in the `Global South' while the companies developing them are based in the `Global North' highlights the digital neocolonialism of AI chatbots \cite{menon2023postcolonial, nyaaba2024generative} -- mimicking the colonial/imperial tradition of using labor and resources from the `Global South' to develop finished products that largely benefit the `Global North'. Such labor is often unfairly compensated and lacks worker protections, as evidenced by OpenAI's practice of outsourcing ChatGPT training to workers in Kenya, Uganda, and India, \cite{perrigo2023exclusive} who earned \$2/hour over 9-hour shifts to label violent, hateful, and sexually explicit content — including detailed descriptions of child sexual abuse, bestiality, murder, and torture, without any support for the emotional and psychological impacts of such work \cite{foxglove2024letter}.  While such `ghost work' in AI training is not a unique phenomenon established to support AI development \cite{gray2019ghost}, the number of model trainers from the `Global South' has only increased, and their working conditions worsened, under the AI chatbot paradigm.  

\subsection{AI Chatbot Development creates Adverse Environmental Effects}\label{sec:environment}

The proliferation of chatbots is also exacerbating the ongoing climate emergency due to the continued development of data centers and server farms required to handle global simultaneous use. While AI systems were already resulting in substantial greenhouse gas emissions before 2023 \cite{dodge2022measuring}, Big Tech companies such as Google and Microsoft have reported significant rises in greenhouse gas emissions since the launch of their chatbots, owing to increased development and usage of data centers \cite{npr2024emissions}. This demonstrates that even though LLMs and multimodal AI models were causing a strain on the environment even before the development and global release of AI chatbots built on top of these models, this phenomenon has significantly worsened the environmental impact.  

Data centers also require millions of gallons of water for round-the-clock cooling \cite{das2023ai, qiao2025you, richie2026reduce}. Data centers can consume 5 million gallons of water per day of operation \cite{halper2023amid}, which is supplied to AI companies managing these data centers at significantly lower rates than those paid by local populations. A planned Google data center in Mesa, Arizona, was found to be paying just over \$6 per 1000 gallons of water, whereas the same volume was costing Mesa residents almost \$11 \cite{sattiraju2020google}, leading to drinking water shortages. A large number of US data centers are also located in regions with moderately to highly stressed watersheds, exacerbating the risk of drinking water shortages and droughts, with significant agricultural impacts \cite{bloomberg2025ai, siddik2021environmental}. Coolant additives used in data centers can also contaminate local water supplies, with Amazon facilities in Morrow County, Oregon, linked to nitrate-based pollutant discharges into drinking water
\cite{cooper2024data, obrien2025amazon}.
Data centers also release high levels of air pollutants such as Nitrogen dioxide, which both cause respiratory issues and contribute to climate change \cite{wa_ecology_datacenters}. 

These concerns are exacerbated in developing countries. In cities such as Visakhapatnam and Bengaluru, which routinely face water shortages, the development of Google and Microsoft data centers has alarmed local populations and environmental advocacy groups alike \cite{imandar2024india, rajesh2024google}. Google paused the development of a \$200 million data center in Santiago, Chile, amid fears and legal pushback that the water consumption from the proposed data center would devastate the city's aquifer at a time when the country is experiencing a nationally-mandated water rationing policy \cite{pbs2024google}. Developing countries thus are having to carefully navigate the difficult choice of how to invest in inviting the development of data centers by enticing global AI companies with mouthwateringly low development costs \cite{livemint2025company} to reap the economic benefits that might be needed for them to keep pace with the rest of the world \cite{dcd_africa_infrastructure}, while also incurring as lightly as possible the environmental and public health costs.

\begin{figure*}[t]
\centering
\resizebox{\textwidth}{!}{%
\hyphenpenalty=10000
\exhyphenpenalty=10000
\setlength{\tabcolsep}{6pt}
\renewcommand{\arraystretch}{1.5}
\scriptsize
\begin{tabular}{L C C C C}
\toprule
\cellcolor{PF} &
\cellcolor{PF}{\color{PT}\footnotesize (N)\par\smallskip\footnotesize\bfseries Non-conversational AI systems} &
\cellcolor{PF}{\color{PT}\footnotesize (M)\par\smallskip\footnotesize\bfseries Modular infrastructure} &
\cellcolor{PF}{\color{PT}\footnotesize (H)\par\smallskip\footnotesize\bfseries Higher-agency chatbot design} &
\cellcolor{PF}{\color{PT}\footnotesize (P)\par\smallskip\footnotesize\bfseries Policy \& institutional safeguards}
\tabularnewline\midrule
\cellcolor{RF}{\bfseries\footnotesize Model layer} &
\cellcolor{SF}{\color{TT}Avoids model-level opacity by design} &
\cellcolor{SF}{\color{TT}Exposes reasoning; auditable components} &
\cellcolor{AF}{\color{AT}Improves transparency; retains LLM opacity} &
\cellcolor{AF}{\color{AT}Can mandate disclosure; limited on internals}
\tabularnewline
\cellcolor{RF}{\bfseries\footnotesize Interface layer} &
\cellcolor{SF}{\color{TT}Sidesteps conversational interface harms entirely} &
\cellcolor{AF}{\color{AT}Legible I/O aids agency; may still use chat front-end} &
\cellcolor{SF}{\color{TT}Directly targets agency, contestability \& harm design} &
\cellcolor{AF}{\color{AT}Content regulation; limited structural change}
\tabularnewline
\cellcolor{RF}{\bfseries\footnotesize Deployment layer} &
\cellcolor{AF}{\color{AT}Reduces infra.\ pressure; varied deployment contexts} &
\cellcolor{SF}{\color{TT}Decentralizes compute; resists vendor lock-in} &
\cellcolor{NF}{\color{GS}Not applicable} &
\cellcolor{SF}{\color{TT}Labor, env.\ \& power safeguards target this layer}
\tabularnewline\bottomrule
\end{tabular}}
\caption{Harms arising from AI systems can be understood across three layers: \textbf{the model layer} (e.g., epistemic risks in LLMs), \textbf{the interface layer} (e.g., agency and relational effects in chatbot interactions), and \textbf{the deployment layer} (e.g., labor, environmental, and power impacts). We propose four complementary intervention strategies: \textbf{(N)} task-specific, non-conversational AI systems, \textbf{(M)} modular AI infrastructure, \textbf{(H)} higher-agency chatbot design, and \textbf{(P)} policy and institutional safeguards. \colorbox{SF}{\color{TT}Green} cells indicate interventions that primarily address a given harm layer; \colorbox{AF}{\color{AT}Yellow} cells indicate partial or contextual coverage.}
\label{fig:layered_harms_interventions}
\end{figure*}

\section{Implications: Resisting AI Paradigm Convergence and Imagining Alternatives}\label{sec:alternatives}

The harms outlined in this paper, across design, interaction, and infrastructure, are not an inevitable endpoint of technical progress.\footnote{See Appendix~\ref{sec:appendix} for a comparison of harm concentrations across AI system types, showing that task-specific and modular systems are largely insulated from the interface-level and relational harms that characterize LLM chatbots.} In this section, we map alternative trajectories that highlight responsibly designed, pluralistic, and agency-preserving AI systems beyond the chatbot paradigm.


\subsection{Stronger Focus on Non-Conversational AI Systems, away from Natural Language Interactions}

Amidst a global focus on developing AI chatbots, there are notable examples of systems straying from this paradigm that share a common philosophy: AI should work for people, not perform intelligence \textit{at} them. One of the strongest examples is in open-source robotics, where low-cost, openly licensed robots and tools built on public platforms are putting AI capabilities directly in the hands of developers, researchers, and educators without a conversational interface mediating that relationship \citeg{huggingfaceReachyMini, techcrunchHuggingFace}. Users build with these tools rather than chat, restoring them as active agents rather than passive recipients of synthesized answers \cite{fortuneStartupPrime}, thus meaningfully achieving the democratization of AI that chatbots seem to strive for \cite{openai_charter}. Features such as specialized interfaces designed around how domain experts think and work, researcher-controlled model infrastructure, and physical automation tools that handle mundane tasks \cite{sfstandardLaundryfoldingRobot} without requiring any prompting all represent a fundamentally different vision of what AI can be. Furthermore, embodied AI systems \citeg{feng2025embodied, zhang2025embodied}, such as AI2's MolmoBOT \cite{deshpande2026molmob0tlargescalesimulationenables}, designed to assist users with real-world tasks, such as grasping and door-opening, could have significant positive impacts on the lives of older adults and individuals with disabilities. These developments pave the way for other sub-fields of AI and suggest that the dominance of the AI chatbot paradigm can be broken, and already, there are developers, researchers, and institutions trying to do so. 

Ultimately, the success of this model of AI development is to expand the range of interaction modalities through which users engage with AI systems, beyond natural language prompts typed into a chatbot interface. Natural language is a powerful and flexible medium, but when positioned as the default or exclusive interface, it can obscure system constraints and collapse complex tradeoffs into seemingly authoritative responses \cite{chow2025beyond}. The conversational interface itself is not neutral: from chat-based layouts to systems using "I" and "me" pronouns, current designs deliberately create the illusion of a conversation partner or a command-executing autonomous entity \cite{bender2026talk}. We need to move beyond the current ``command-based'' interactions to ``intent-based'' interactions, thus providing more clarity on the shared control aspect of human-AI interaction \cite{kraljic2024prompt}. Alternative modalities could include visual parameter spaces, interactive simulations, rule-based editors, sliders, and structured query interfaces. 
Recent industry products gesture in this direction even within firms whose
primary offerings are conversational: Anthropic's Claude Design grounds
editing in direct manipulation of a generated canvas \cite{anthropic2026claudedesign}, and Google's NotebookLM organizes
interaction around user-uploaded sources with inline citations rather than
open-ended chat \cite{google2026notebooklm}.
These forms of interaction allow users to explore how outputs change in response to inputs, to inspect uncertainty, and to develop an intuitive understanding of system behavior without the misleading social cues of conversation. Natural language can remain available as a supplementary modality without monopolizing interaction or suggesting the presence of an interlocutor. This pathway positions interface design as a site of agency restoration: by making system behavior legible and manipulable through non-conversational means, multimodal interfaces enable users to reason with AI systems rather than simply defer to them. Moving beyond natural language as the dominant modality also requires moving beyond anthropomorphic framing that treats systems as agents who ``help," ``create," or ``collaborate," toward language that accurately describes users employing computational tools for specific purposes.

\subsection{Designing AI as (Modular) Infrastructure}

Upending the AI chatbot paradigm requires a fundamental shift in thinking, potentially through a reorientation of priorities of AI systems toward infrastructural integration. Research showing that AI systems secretly rely on vast amounts of hidden human labor undermines the claim that chatbot interfaces represent true machine intelligence \cite{nemer2025artificial}. AI systems do not actually run on machine intelligence alone; they lean on the labor of human workers who label data, moderate content, and make judgment calls that the system passes off as its own. Designing AI as infrastructure makes this dependence legible by foregrounding the social and material conditions that allow for intelligent systems (Section \ref{subsec:labor}) instead of obscuring them behind the illusion of autonomous dialogue. In this approach, AI doesn't pretend to be someone being spoken to. Instead, it sits quietly inside existing tools and workflows, takes defined inputs, and produces outputs users can inspect and verify. Reliability and transparency matter more than sounding fluent and human.

Designing AI as infrastructure foregrounds reliability, interpretability, and task alignment over linguistic fluency. This spreads AI capability across many specific, user-controlled tools instead of funneling everything through one centralized conversational agent. First, it restores user agency. When AI is a specific tool that users deliberately invoke for a specific purpose, they remain in control, decide what inputs to give it, can inspect its outputs, and are not being quietly funneled toward a single authoritative-sounding answer with no alternatives offered. Next task-specific tools (ideally) require users to still do the surrounding work, like formulating the problem, interpreting results, and making judgments. Thus, users rely on AI for a defined subtask rather than handing the whole problem over to a chatbot, which keeps reasoning skills intact rather than gradually offloading them to a system that does the thinking for them.

Furthermore, architectural diversification through modularity and composability allows AI capabilities to be decomposed into interoperable components with explicit inputs/outputs, and responsibilities. Users, developers, or institutions can then assemble these into pipelines that reflect local goals, values, and constraints. When systems offer recommendations, users should be able to inspect the data that informed them, the alternatives evaluated, the associated confidence or uncertainty, and the sources available for verification. The aim is to make system behavior intelligible and open to challenge, preserving understanding of how/why outputs are produced \cite{raees2024explainable}. Modular architectures \citeg{substackComfyUIR1ComfyUI} offer several advantages beyond technical flexibility. They reduce dependency on single vendors, enable targeted auditing and evaluation, and make it easier to replace or remove components that perform poorly or embody misaligned assumptions. Composability also supports experimentation: different models or methods can be swapped in without requiring wholesale adoption of a new paradigm. 

\subsection{Designing Higher-Agency AI Chatbots}

Even within the AI chatbot paradigm, there is room for improvement in terms of developing chatbots that prioritize and maintain user agency in their interactions. For instance, instead of offering a single answer per prompt (Section \ref{subsec:chatbot-design}), chatbots can be designed to provide multiple different perspectives or solutions to subjective (descriptive or provocative queries) or semi-subjective questions (e.g., coding solutions with multiple correct approaches). Researchers such as \citet{wang2022self} have explored such approaches through model self-consistency, which can be implemented in how chatbots deliver their outputs. Furthermore, the field of Explainable AI has long since advocated for the accompaniment of explanations -- such as chains-of-thought, primary sources considered, and other such factors -- alongside AI ``decisions,'' showing how the presence of explanations helps users better understand outputs and diagnose errors \cite{cohen2024contextcite}. Chatbot developers should also consider tighter protections around the production of harmful content, such as deepfakes/cheapfakes and non-consensual intimate imagery (Section \ref{subsec:new-harm}), attempting to identify user intent in creating such content and implementing early prevention by generating black-squared images \citeg{ghosh2023person} or other means. 

\subsection{Policy Considerations}\label{subsec:policy}

Any meaningful shift of the field of AI development away from the chatbot paradigm cannot be achieved through technical perspectives alone, and will require significant institutional support and policy intervention geared towards a stronger and more collective focus on reducing ``the capitalist, imperialist and environmental dimensions of digital power." \cite{kwet2025digital} Public procurement policies can incentivize task-specific, agency-preserving AI systems by requiring transparency, interpretability, and workflow integration in government, educational, and healthcare contexts. Funding mechanisms should explicitly support pluralistic system design with open, efficient models that can be built on by communities rather than rewarding scale and generality. Labor protections must address the erosion of entry-level pathways to expertise, exploitation of crowdworkers in AI training pipelines, and devaluation of creative work. Environmental regulations should require impact and rising cost assessments (like those mentioned in Section \ref{subsec:economy} and \ref{sec:environment}) for large-scale AI deployments, water use restrictions in drought-prone regions, fair resource pricing that internalizes currently externalized costs, and incentives for building more environmentally friendly data centers, such as Vietnam's Viettel Hoa Lac Data Center, which uses renewable energy to meet 30\% of its electricity needs \cite{viettel2024green}. 

Beyond procurement, labor, and environmental measures, paradigm diversification requires attention to the ownership and governance structures through which AI development currently proceeds. Modular and task-specific systems built on the same concentrated compute, data, and labor arrangements as current chatbot infrastructure will reproduce those arrangements' harms regardless of interface design. Structural change therefore requires public investment in alternative compute infrastructure that is not dependent on a small number of vendors, enforceable labor protections across training pipelines including the demands articulated by organizations such as the African Content Moderators Union and legal advocates such as Foxglove, and community governance over data center siting and water and energy allocation of the kind exercised in successful challenges to projects including Google's proposed Santiago facility \cite{pbs2024google}. These arrangements treat affected communities as sources of analysis and as participants in governance rather than as sites where harms accumulate or as audiences for technical fixes.

Above all, addressing the concerns outlined in Section \ref{sec:societal} requires a field-wide understanding that AI chatbots do not improve every (or even the majority of) situation(s) in which they are deployed \citeg{humlum2025large}, and may actually create worse outcomes than non-AI-assisted solutions \citeg{green2020algorithmic}. Recognizing this is a precondition for developing systems that are more closely aligned with actual user needs (Section \ref{subsec:user-needs}).

\section{Limitations and Future Work}

Our analysis focuses primarily on general-purpose conversational AI systems deployed for broad consumer and professional use, deliberately excluding specialized agents with narrow domains and domain-specific AI assistants that preserve traditional workflows. This scoping decision, while necessary for analytical clarity, means our critique may not fully capture the heterogeneity of conversational AI implementations or contexts in which conversational interfaces may be genuinely appropriate. Much of our evidence draws from emerging research on relatively recent systems, and longer-term effects of widespread chatbot adoption on cognitive practices, social relations, and professional expertise remain to be studied. While we acknowledge that conversational interfaces offer genuine benefits in specific contexts (particularly for users with certain disabilities, language learners, or individuals facing barriers to traditional computing interfaces), our critique targets the universalization and monopolization of this paradigm, not the existence of conversational AI as one option among many. 

Empirical work is needed to document the long-term cognitive, social, and professional effects of sustained chatbot use across different domains and populations. Design research should explore and evaluate the alternative interaction paradigms we have proposed, prototyping workflow-preserving AI systems, making modular architectures accessible to non-experts, and identifying interface patterns that support transparency and contestability while remaining usable. Critical attention should be directed towards measuring the global and distributional dimensions of AI paradigm convergence, examining how costs and benefits differ across regions, socioeconomic positions, and cultural contexts. Theoretical work is needed to refine frameworks that distinguish between forms of automation that enhance human capability and those that diminish it. Addressing these requires sustained interdisciplinary collaboration among researchers, designers, policymakers, and affected communities.

\section{Conclusion}


In this paper, we argue that the rapid convergence toward conversational AI chatbots does not represent a neutral technological trajectory, but a consequential reconfiguration of how humans interact with information, institutions, and one another. The chatbot paradigm systematically erodes user agency while appearing to enhance it, reshapes cognitive practices in ways that prioritize convenience over deliberation, and imposes substantial environmental and economic costs that fall disproportionately on marginalized populations. This paradigm is neither inevitable nor irreversible, as outlined alternative pathways demonstrate that other forms of AI development remain not only technically feasible but may even be socially preferable. The question is not whether AI will shape future human activity, but which forms of AI will do so and under what conditions. Do we want a single interaction paradigm to dominate because it aligns with narrow definitions of economic value, or will we deliberately cultivate a pluralistic ecosystem that reflects diverse needs, contexts, and views? The answer will determine not only what AI looks like, but what human agency, expertise, and care come to mean in an increasingly automated world.

\begin{acks}
The authors are grateful to Margaret Mitchell and Stella Biderman for their constructive comments during the manuscript preparation.
\end{acks}

\section*{Author Positionality}

All four authors are presently affiliated with academic institutions and technology research organizations in the United States, while having grown up in the Global South. We come from varied relationships to the systems we critique, spanning industry research, distributed AI development, and academic work centered on historically marginalized communities. Our backgrounds and current environments inform our attention to how the chatbot paradigm concentrates power unevenly across geographies, labor markets, and populations.

\section*{Generative AI Disclosure Statement}

We used Generative AI (Claude 4.6 and Gemini 3.0) for generic conversational brainstorming, followed by manual paper writing. Upon completion of the manuscript, we used Grammarly AI, ChatGPT, and Claude for grammar corrections and sentence restructuring, and an Agentic Reviewer (\url{https://paperreview.ai/}) to iteratively gather feedback and improve the paper. 

\newpage
\bibliographystyle{ACM-Reference-Format}
\bibliography{sample-base}

\section{Appendix}
\label{sec:appendix}
 
Table~\ref{tab:ai_harms_example} maps harm categories onto five classes of AI systems that differ along model, interface, and deployment dimensions. This complements Figure~\ref{fig:layered_harms_interventions}, which maps intervention strategies onto harm layers: where Figure~\ref{fig:layered_harms_interventions} asks \textit{what can be done}, this table asks \textit{which systems are implicated}. We use it to motivate our focus on LLM-based chatbots as the system class that uniquely concentrates harms across all categories.
 
\textbf{General AI} refers to the broad class of machine learning and automated decision systems deployed across domains, not limited to language or conversation.
\textbf{LLMs} are generative language models that produce or transform text, but are not necessarily deployed through a conversational interface.
\textbf{Chatbots} are conversational interfaces that enable interaction through dialogue, independent of the underlying model, including rule-based and retrieval-based systems.
\textbf{LLM Chatbots} are general-purpose conversational systems built on LLMs and are the primary focus of this paper.
\textbf{Task-specific AI} refers to non-conversational machine learning systems embedded in structured workflows with narrow, domain-specific functionality (e.g., AlphaFold, FourCastNet).
 
The table shows that while individual harm categories are distributed across system types, LLM-based chatbots are the only class that is primarily or strongly associated with harms across \textit{all} categories. Task-specific AI, by contrast, is largely insulated from interface-level and relational harms, reinforcing our argument that diversifying beyond the dominant chatbot paradigm, whether through task-specific systems, modular infrastructure, higher-agency chatbot design, or policy safeguards, can meaningfully reduce harm concentration.
 
\begin{table*}[htpb!]
\centering
\footnotesize
\begin{tabular}{p{4.5cm} c c c c c}
\hline
\textbf{Harm type} &
\textbf{General AI} &
\textbf{LLMs} &
\textbf{Chatbots} &
\textbf{LLM chatbots} &
\textbf{Task-specific AI} \\
\hline
 
Cognitive offloading \& skill loss &
\yes & \yes & \partialyes & \yes & \partialyes \\
 
Epistemic (hallucination, misinformation) &
\partialyes & \yes & \no & \yes & \partialyes \\
 
Agency \& autonomy reduction &
\partialyes & \partialyes & \partialyes & \yes & \no \\
 
Relational \& social harms &
\no & \partialyes & \partialyes & \yes & \no \\
 
Novel harm vectors (deepfakes, NCII) &
\no & \yes & \no & \yes & \no \\
 
Economic \& labor harms &
\yes & \partialyes & \partialyes & \yes & \partialyes \\
 
Power centralization \& lock-in &
\yes & \partialyes & \no & \yes & \partialyes \\
 
Environmental \& infrastructure harms &
\yes & \yes & \no & \yes & \partialyes \\
 
Contestability \& transparency challenges &
\yes & \yes & \partialyes & \yes & \partialyes \\
 
\hline
\end{tabular}
 
\vspace{2mm}
\textit{\yes\ = primary / strongly associated;\quad
\no\ = largely not applicable;\quad
\partialyes\ = partially or contextually applicable.}
 
\caption{Harms across different AI system types. Rows represent harm categories, while columns distinguish classes of AI systems that differ along model, interface, and deployment dimensions. LLM-based chatbots are the only system class strongly associated with harms across all categories.}
\label{tab:ai_harms_example}
\end{table*}

\end{document}